\documentclass[journal=jacsat,manuscript=article]{achemso}
\usepackage{chemformula} % Formula subscripts using \ch{}
\usepackage[T1]{fontenc} % Use modern font encodings
\usepackage{amssymb}
\usepackage{float}
\usepackage[section]{placeins}

\DeclareUnicodeCharacter{2212}{\ensuremath{-}}

\title{Weak Host Interactions Induced Thermal Transport Properties of Metal Halide Perovskites Deviating from the Rattling Model}
\author{Yu Wu}
\email{wuyu@csj.uestc.edu.cn}
\affiliation{Yangtze Delta Region Institute (Huzhou), University of Electronic Science and Technology, Huzhou, Zhejiang 313001, China}

\author{Linxuan Ji}
\affiliation{School of Physics, State Key Laboratory of Electronic Thin Films and Integrated Devices, University of Electronic Science and Technology, Sichuan, Chengdu 611731, China}

\author{Shuming Zeng}
\affiliation{College of Physics Science and Technology, Yangzhou University, Jiangsu 225009, China}

\author{Yimin Ding}
\affiliation{Yangtze Delta Region Institute (Huzhou), University of Electronic Science and Technology, Huzhou, Zhejiang 313001, China}

\author{Liujiang Zhou}
\email{ljzhou@uestc.edu.cn}
\affiliation{School of Physics, State Key Laboratory of Electronic Thin Films and Integrated Devices, University of Electronic Science and Technology, Sichuan, Chengdu 611731, China}
\alsoaffiliation{Yangtze Delta Region Institute (Huzhou), University of Electronic Science and Technology, Huzhou, Zhejiang 313001, China}

\begin{document}

\begin{abstract}

The low-frequency phonon branches of metal halide perovskites typically exhibit the characteristic of hardening with the increase of the cation mass, which leads to anomalous thermal transport phenomenon. However, the underlying physical mechanism is not yet understood. Here, we theoretically compare the thermal transport properties of $A_2$SnI$_6$ ($A$=K, Rb, and Cs) perovskites. The thermal transport in perovskites is widely explained using the rattling model, where ``guest'' cations inside the metal halide framework act as ``rattlers'', but this fails to explain the following phenomenon: The low-frequency phonon branch of $A_2$SnI$_6$ perovskites is insensitive to the mass of the $A^+$ cation and strongly correlated with the interaction of the $A^+$ cation with the I$^-$ anion in the octahedral structures. %The strongest $A$-I interaction in Cs$_2$SnI$_6$ results in the largest particlelike thermal conductivity of Cs$_2$SnI$_6$. 
The failure of the rattling model stems mainly from the weak interactions between the octahedral structures. And we find that this weak interaction is prevalent in other metal halide perovskites. By developing a new spring model, we successfully explain the thermal transport behavior in $A_2$SnI$_6$ perovskites. Our work gives new insights into the thermal transport mechanism in metal halide perovskites, which has a guiding significance for designing extremely low thermal conductivity materials.

%Suppressed three-phonon scattering processes have been considered to be the direct cause of materials exhibiting significant higher-order four-phonon interactions. However, after calculating the phonon-phonon interactions of 128 Half-Heusler materials by high-throughput, we find that the acoustic phonon bandwidth dominates the three-phonon and four-phonon scattering channels and keeps them roughly in a co-increasing or decreasing behavior. The $aao$ and $aaa$ three-phonon scattering channels in Half-Heusler materials are weakly affected by the acoustic-optical gap and acoustic bunched features respectively only when acoustic phonon bandwidths are close. Finally, we found that Half-Heusler materials with smaller acoustic bandwidths tend to have a more pronounced four-phonon effect, although three-phonon scattering may not be significantly suppressed at this time. \textcolor{black}{Our work gives new insights into the high-order four-phonon effect in solids, which has guiding significance for designing thermal management materials.}

\end{abstract}

\flushbottom
\maketitle

\thispagestyle{empty}

\section*{Introduction}

The metal halide perovskites are widely used in the field of photovoltaics\cite{Jung2019,Saliba2016}, light-emitting diodes\cite{Wang2019a,Lim2019}, photodetectors\cite{Yang2018a,Pan2017,Xiang2019}, and thermoelectrics\cite{Zeng2022a,Sajjad2020} owing to their excellent light absorption capability, the relatively long diffusion length of charge carriers and tunable bandgap. Understanding the heat transport mechanism of perovskites is essential for the thermal management of perovskite-based devices. On the one hand, devices such as photodetectors require materials with high thermal conductivity to provide heat dissipation and thus enhance operational lifetime. On the other hand, the thermoelectric field requires low lattice thermal conductivities for large energy conversion efficiencies.

The heat transport mechanism of perovskites is closely related to its special crystal structure. Take the common $ABX_3$-type metal halide perovskites as an example, the framework is composed of corner-shared $BX_3^-$ octahedra with metals (e.g., Pb$^{2+}$, Sn$^{2+}$) at the center and halide anions at the
corners. The $A$-site cations (e.g., Cs$^+$, organic cations) are located in the hollow within the perovskite framework. The soft phonon modes and large lattice anharmonicity results from the rattling motion of $A$-site cations and dynamic rotation of halide atoms have long been considered to be the key factors limiting lattice thermal transport of metal halide perovskites\cite{Bhui2022,Wu2024,Acharyya2020,Wu2024a}. However, these theories cannot account for the following facts: 1) There is no distinct avoid-crossing phenomenon between the acoustic and optical branches\cite{Wang2023,Zeng2022a}, which is a typical characteristic of the rattling model\cite{Christensen2008,Dutta2020,Lin2016,Li2022a}. 2) As the mass of $A$-site cations increases, the low-frequency phonon branches harden, which further brings about anomalous mass-dependent thermal transport phenomenon\cite{Varadwaj2020,Mahmood2023,Cheng2024,Ayyaz2024}. Moreover, the thermal transport of the CsPbI$_3$ and empty PbI$_6$ frameworks was compared and it was found that Cs$^+$ cations lead to an enhancement in thermal conductivity\cite{Thakur2023}. Therefore, exploring the effect of $A$-site cations and the failure of the rattling model is critical to understanding the thermal transport properties of metal halide perovskites.

Here, we studied the thermal transport properties of a class of vacancy-ordered metal halide perovskites $A_2$SnI$_6$ ($A$=K, Rb, and Cs). The structure can be regarded as a defect variant of $A$SnI$_3$ with isolated [SnI$_6$]$^{2-}$ octahedra bridged by $A$-site cations. The results show that Cs-based perovskites have greater $A$-I interactions. The low-frequency phonon branches are not sensitive to the mass of $A$-site cations and strongly correlate with the $A$-I interactions, which leads to a larger particlelike thermal conductivity of Cs$_2$SnI$_6$. The rattling model failure arises from weak interactions between the octahedra, which is inconsistent with the strong interaction condition of the host frame in the model. By building a spring model with weak host interaction, we successfully explain the anomalous thermal transport behavior of $A_2$SnI$_6$. Further analysis shows that this weak interaction still exists even though the octahedral structure is corner-shared, making this anomalous behavior prevalent in metal halide perovskites.

\section*{Results and Discussion}

\begin{figure*}[ht!]
\centering
\includegraphics[width=1\linewidth]{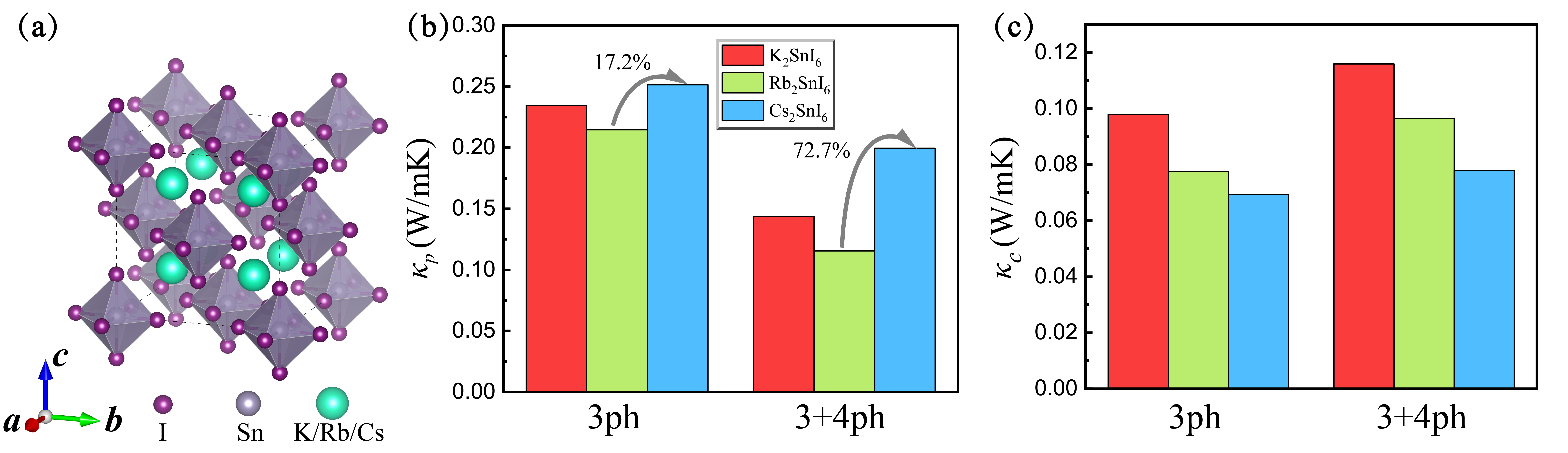}
\caption{(a) The conventional cell of $A_2$SnI$_6$ perovskites. (b) The particlelike thermal conductivities ($\kappa_p$) of $A_2$SnI$_6$ perovskites at 300 K. (c) The glasslike thermal conductivities ($\kappa_c$) of $A_2$SnI$_6$ perovskites at 300 K.}
\label{Fig1}
\end{figure*}

The $A_2$SnI$_6$ ($A$=K, Rb, and Cs) perovskites have a face-centered cubic lattice with space group $Fm\overline{3}m$ and are typically described as $A^{+}_2$Sn$^{4+}$I$^{-}_6$.  Each Sn$^{4+}$ is bonded to six equivalent I$^-$ to form an isolated [SnI$_6$]$^{2-}$ octahedron, which is bridged by $A^+$ cations as shown in Fig.~\ref{Fig1}(a). The structure of $A_2$SnI$_6$ perovskites differs from the host-guest system corresponding to the rattling model, where the adjacent atoms in the host have bonding connections. Specifically, the $A^+$ cations can be viewed as guest atoms, however, the [SnI$_6$]$^{2-}$ octahedra cannot be viewed as host frames in the traditional sense due to the lack of strong bonding between them. Figure~\ref{Fig1}(b) shows the particlelike thermal conductivities ($\kappa_p$) of $A_2$SnI$_6$ perovskites at 300 K. $A_2$SnI$_6$ perovskites exhibits anomalous mass-dependent $\kappa_p$. The Cs$_2$SnI$_6$ perovskite has the largest total atomic mass but exhibits the highest $\kappa_p$. Considering three-phonon (3ph) scattering, The $\kappa_p$ of Cs$_2$SnI$_6$ is 17.2\% higher than that of Rb$_2$SnI$_6$. When four-phonon (4ph) scattering is introduced, the difference between $\kappa_p$ is further amplified to 72.7\%. The glasslike thermal conductivity ($\kappa_c$) contributed by wavelike phonon tunneling in $A_2$SnI$_6$ perovskites cannot be ignored as shown in Fig.~\ref{Fig1}(c). Taking Cs$_2$SnI$_6$ as an example, $\kappa_c$ occupies 28.0\% of the total thermal conductivity ($\kappa_L=\kappa_p+\kappa_c$). The $\kappa_c$ of $A_2$SnI$_6$ perovskites exhibits two distinctive features: 1) It shows a gradual decrease with total atomic mass. 2) Considering 3+4ph scattering, $\kappa_c$ is significantly improved compared to the result when considering only 3ph scattering. $\kappa_c$ is closely related to phonon lifetime, and low lifetime phonons will contribute more to $\kappa_c$\cite{DiLucente2023,Ji2024,Wu2024,Tong2023,Wu2023b}.

\begin{figure*}[ht!]
\centering
\includegraphics[width=1\linewidth]{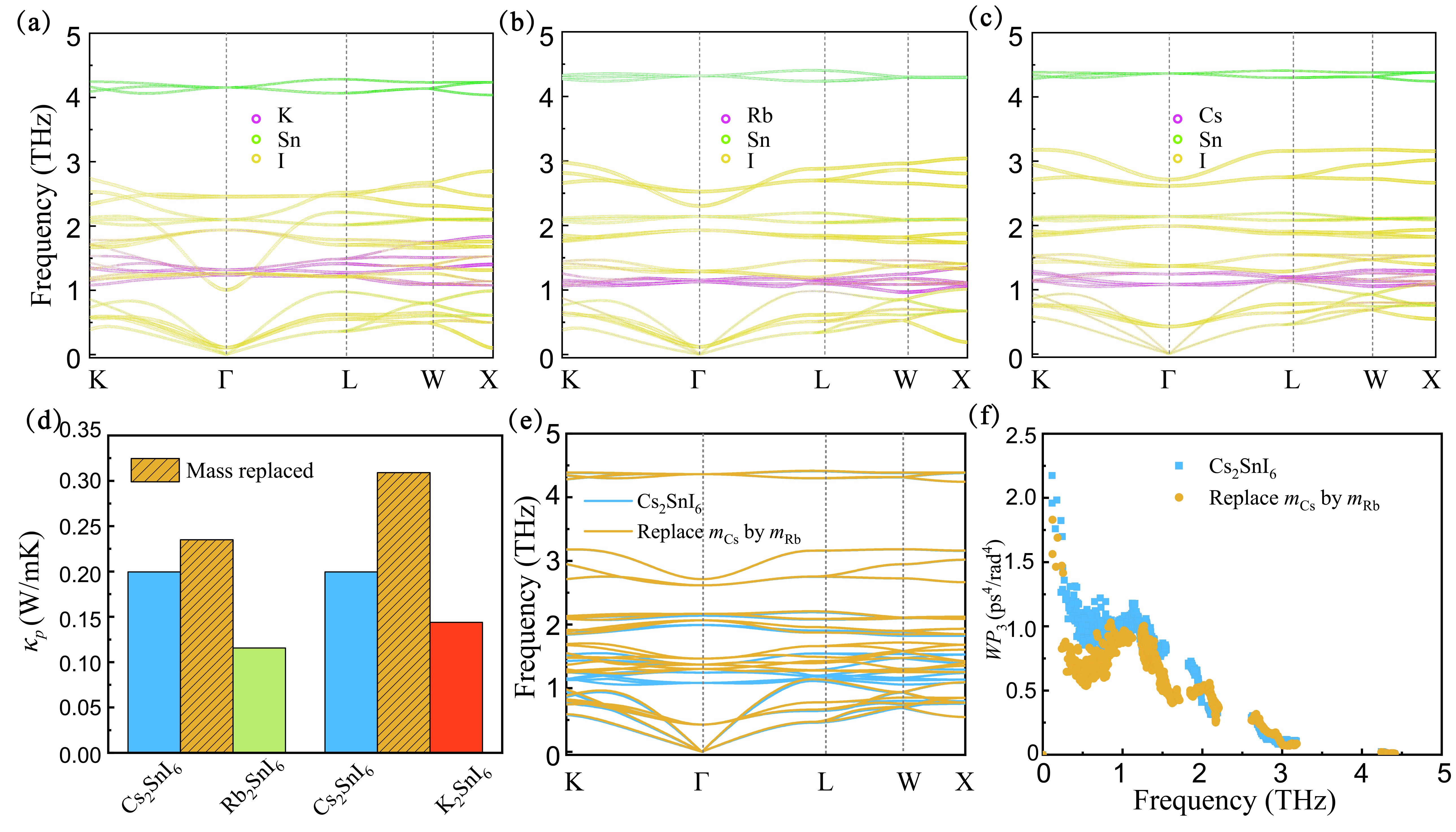}
\caption{(a-c) The phonon dispersion projected by the eigenvectors of each atom in $A_2$SnI$_6$ perovskites. (d) The $\kappa_p$ after modifying the atomic mass of Cs in Cs$_2$SnI$_6$. (e) The phonon dispersion and (f) $WP_3$ of the original Cs$_2$SnI$_6$ and the phonon dispersion after replacing the Cs mass with the Rb mass.}
\label{Fig2}
\end{figure*}

Figure~\ref{Fig2}(a-c) shows the phonon dispersion projected by the eigenvectors of each atom. The low-frequency (<1 THz) phonon branches are mainly contributed by the I$^-$ in the octahedral structure. From K$_2$SnI$_6$ to Cs$_2$SnI$_6$, the low-frequency phonon branches gradually harden. Especially for Cs$_2$SnI$_6$, the frequency of the first optical branch at $\Gamma$ point is 3.6 times that of Rb$_2$SnI$_6$. The $A^+$ cations mainly contribute to optical branches in the range of 1 to 1.5 THz in $A_2$SnI$_6$. It is noted that there is a clear boundary between the optical branches contributed by the $A^+$ cations and the acoustic branches, which is inconsistent with the characterization of the rattling atom. In the classical rattling model, the low-frequency optical phonons contributed by guest atoms and the acoustic phonons are coupled together and induce the avoid-crossing feature of phonon dispersion. To investigate the effect of the atomic mass of the $A$-site on the thermal transport of $A_2$SnI$_6$ perovskites, Fig.~\ref{Fig2}(d) shows the change of $\kappa_p$ after modifying the atomic mass of Cs in Cs$_2$SnI$_6$. Replacing the mass of Cs with
K and Rb in Cs$_2$SnI$_6$ results in a further increase in $\kappa_p$. The phonon dispersion of the original Cs$_2$SnI$_6$ and the phonon dispersion after replacing the Cs mass with the Rb mass are shown in Fig.~\ref{Fig2}(e). The $A$-site atomic mass has a negligible effect on the low-frequency phonon branch at frequencies smaller than 1 THz, and the decrease in the $A$-site atomic mass causes the optical branch it contributes to move toward higher frequencies. This further reduces the scattering channels of acoustic phonons as can be seen from the weighted 3ph scattering phase space ($WP_3$) in Fig.~\ref{Fig2}(f). The dependence of the phonon dispersion on $A^+$ mass is markedly different from the typical material YbFe$_4$Sb$_{12}$ for which the rattling model applies as seen in Fig.~S1.

\begin{figure*}[ht!]
\centering
\includegraphics[width=1\linewidth]{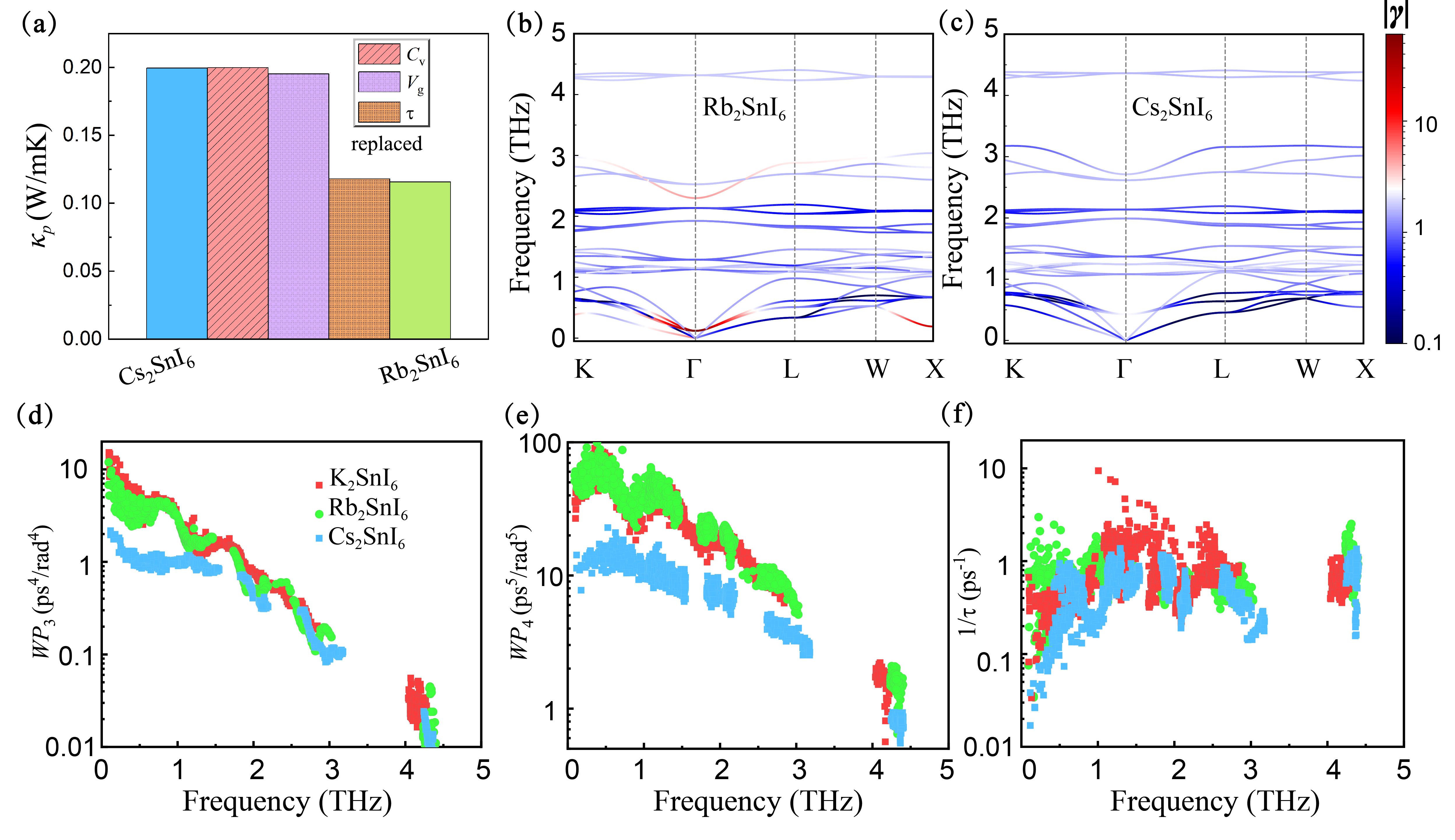}
\caption{(a) The calculated $\kappa_p$ at 300 K and the cross-calculated $\kappa_p$ with the independent replacement of heat capacity, group velocity, and phonon lifetime. The phonon dispersion of (b) Rb$_2$SnI$_6$ and (c) Cs$_2$SnI$_6$ projected by $|\gamma|$. (d) 3ph and (e) 4ph weighted phase space of $A_2$SnI$_6$ perovskites. (f) The scattering rates of $A_2$SnI$_6$ perovskites.}
\label{Fig3}
\end{figure*}

To understand the origin of the $\kappa_p$ enhancement of Cs$_2$SnI$_6$ perovskite, we examined the individual contributions of the heat capacity $C_v$, group velocity $v_g$, and phonon lifetime $\tau$ by replacing mode-resolved $C_v$, $v_g$, and $\tau$ of Cs$_2$SnI$_6$ independently with those of Rb$_2$SnI$_6$\cite{Zhang2023b}. Replacing $C_v$ and $v_g$ has a negligible effect on $\kappa_p$. In contrast, the $\kappa_p$ drops dramatically after the replacement $\tau$, which is close to the result of Rb$_2$SnI$_6$. The $\tau$ is closely related to the scattering induced by phonon-phonon interactions, and the scattering strength and the scattering channels determine its magnitude. The scattering strength is related to the lattice anharmonicity, which can be reflected by the Gr{\"u}neisen parameter $\gamma$. Figure~\ref{Fig3}(b, c) shows the phonon dispersion of Rb$_2$SnI$_6$ and Cs$_2$SnI$_6$ projected by $|\gamma|$. The $|\gamma|$ in Rb$_2$SnI$_6$ is generally higher than that in Cs$_2$SnI$_6$. At the $\Gamma$ point, the $|\gamma|$ of the first optical branch in Rb$_2$SnI$_6$ is an order of magnitude higher than that in Cs$_2$SnI$_6$. The scattering channels can be reflected by weighted phase space. Figure~\ref{Fig3}(d, e) shows the 3ph ($WP_3$) and 4ph ($WP_4$) weighted phase space of $A_2$SnI$_6$ perovskites, respectively. Cs$_2$SnI$_6$ has significantly smaller 3ph and 4ph scattering channels below 1 THz compared to K$_2$SnI$_6$ and Rb$_2$SnI$_6$ due to the hardening of low-frequency phonons, and phonons in this frequency range make a major contribution to $\kappa_p$ as seen in Fig.~S2. Figure~\ref{Fig3}(f) shows the phonon scattering rate, which is the reciprocal of the lifetime (1/$\tau$). The weak phonon scattering strength and scattering channels in Cs$_2$SnI$_6$ make it the lowest scattering rate among the $A_2$SnI$_6$ perovskites.

\begin{figure*}[ht!]
\centering
\includegraphics[width=1\linewidth]{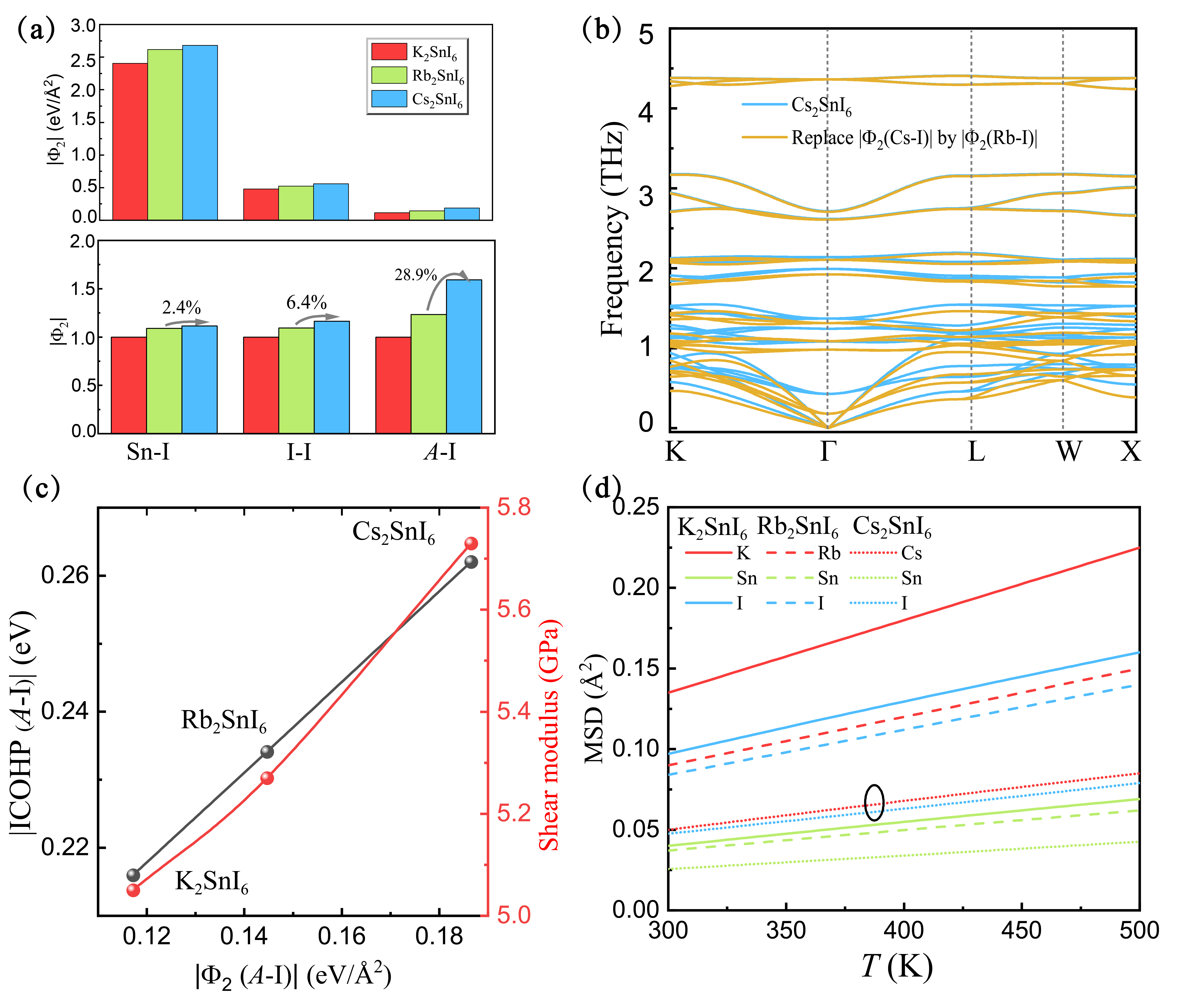}
\caption{(a) The norm of the second-order force constants $|\Phi_2|$ for the three strongest bonds in $A_2$SnI$_6$ perovskites. The bottom shows the results with $|\Phi_2|$ in K$_2$SnI$_6$ being shifted to unity. (b) The phonon dispersion of Cs$_2$SnI$_6$ and that after replacing $|\Phi_2(\rm{Cs-I})|$ by $|\Phi_2(\rm{Rb-I})|$. (c) The |ICOHP| value at the Fermi level and shear modulus as a function of the $|\Phi_2|$. (d) Calculated temperature-dependent atomic mean square displacements (MSD) of $A_2$SnI$_6$ perovskites.}
\label{Fig4}
\end{figure*}

To investigate the origin of the differences in the phonon lifetimes of $A_2$SnI$_6$ perovskites, the norm of the second-order force constants $|\Phi_2|$ for the three strongest bonds are shown in the top of Fig.~\ref{Fig4}(a). The bottom shows the results with $|\Phi_2|$ in K$_2$SnI$_6$ being shifted to unity. In contrast to a relatively small increase for Sn-I and I-I bonds, the $|\Phi_2|$ of $A$-I bond in Cs$_2$SnI$_6$ increase by 28.9\% compared with Rb$_2$SnI$_6$. Figure~\ref{Fig4}(b) shows the phonon dispersion of Cs$_2$SnI$_6$ and that after replacing $|\Phi_2(\rm{Cs-I})|$ by $|\Phi_2(\rm{Rb-I})|$. Significant softening of the low-frequency phonon branches occurs after modification of the $|\Phi_2|$, approaching the results of Rb$_2$SnI$_6$. The change in $|\Phi_2|$ can be related to the electronic bonding via the integrated crystal orbital Hamiltonian population (ICOHP)\cite{Dronskowski1993}. As seen in Fig.~\ref{Fig4}(c), the |ICOHP| value at the Fermi level for $A$-I bond matches the trend of the corresponding $|\Phi_2|$. At the same time, the strong $A$-I bond increases the rigidity of the material which the shear modulus can reflect. Further, it can be seen in Fig.~\ref{Fig4}(d) that the strong $A$-I bond in Cs$_2$SnI$_6$ greatly weakens the mean square displacements (MSD) of the Cs and I atoms.

\begin{figure*}[ht!]
\centering
\includegraphics[width=1\linewidth]{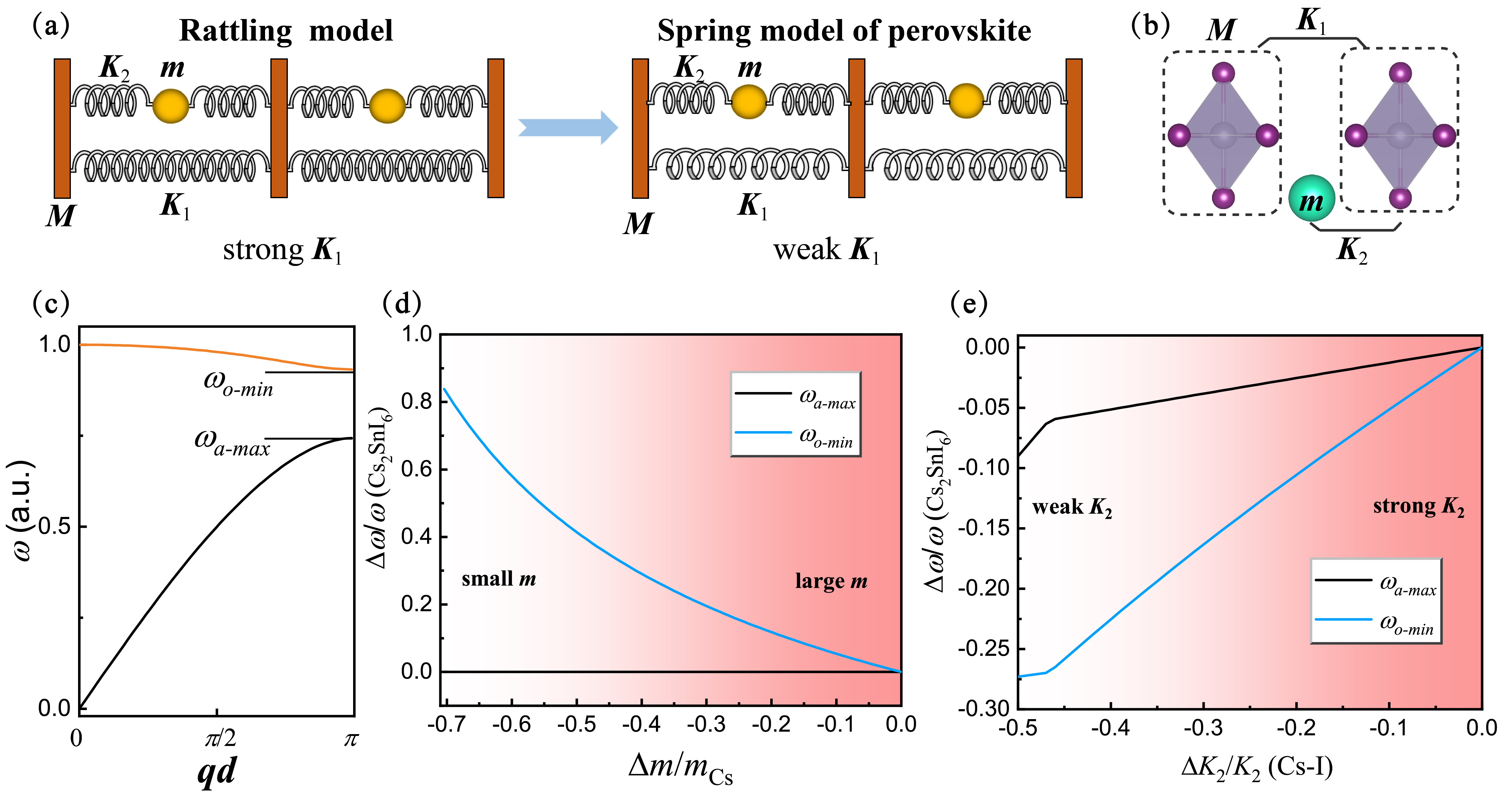}
\caption{(a) Comparison between rattling model and spring model of $A_2$SnI$_6$ perovskites. (b) The structural units in the $A_2$SnI$_6$ perovskites coincide with the established spring model. (c) The analytical solution of the spring model corresponds to Cs$_2$SnI$_6$. The variation rate of $\omega_{a-max}$ and $\omega_{o-min}$ after varying (d) $m$ as well as (e) $K_2$ on the basis of the Cs$_2$SnI$_6$ spring model parameters.}
\label{Fig5}
\end{figure*}

The above discussion reveals an important feature of the thermal transport of $A_2$SnI$_6$ perovskites: The low-frequency phonon branches, which are mainly contributed by the [SnI$_6$]$^{2-}$ octahedra, are not sensitive to the mass of $A^+$ cations, but are sensitive to the interaction between $A^+$ and I$^-$ in [SnI$_6$]$^{2-}$ octahedra. Moreover, the disappearing avoid-crossing phenomenon of phonon dispersion suggests that the classical rattling model cannot explain the thermal transport properties. Based on the characterization of the weak interaction of the octahedra in $A_2$SnI$_6$ perovskites, we developed a spring model applicable to its thermal transport. As seen in Fig.~\ref{Fig5}(a), compared to the classical rattling model\cite{Christensen2008}, this model has a weak elastic constant $K_1$ between host frames. Figure~\ref{Fig5}(b) illustrates the structural units in the $A_2$SnI$_6$ perovskites that coincide with the established spring models. The mass of atom $A$ is $m$. See [SnI$_6$]$^{2-}$ as a whole, where the sum of the mass of all atoms is $M$. The elastic constants $K_1$ and $K_2$ are determined by the $\Phi_2$ between atoms. The parameters used in the spring model are shown in Table S1. Figure~\ref{Fig5}(c) shows the analytical solution of the spring model corresponding to Cs$_2$SnI$_6$. The acoustic branch frequency maximum and the optical branch frequency minimum are defined as $\omega_{a-max}$ and $\omega_{o-min}$, respectively. Figures~\ref{Fig5}(d, e) give the variation of $\omega_{a-max}$ and $\omega_{o-min}$ after varying $m$ as well as $K_2$ on the basis of the Cs$_2$SnI$_6$ spring model parameters, respectively. $\omega_{a-max}$ is insensitive to changes in $m$, and the gradual increase with $K_2$ is consistent with the rule demonstrated in Fig.~\ref{Fig2}(e) and Fig.~\ref{Fig4}(b). To investigate the effect of $K_1$ on phonon dispersion, Fig.~S3(a) shows the solution of the spring model of Cs$_2$SnI$_6$ with 5 times the original $K_1$. The avoid-crossing phenomenon occurs between the acoustic and optical branches, returning to the classical rattling model. At this time, with the increase of $m$, $\omega_{a-max}$ shows a gradual downward trend in general.

It is a common phenomenon that phonons of Cs-based halide perovskites are harder than Rb-based halide perovskites. Although the two neighboring octahedra in our chosen halide chalcogenide system do not share halide atoms, the spring model we have developed is equally applicable to the case of shared halide atoms. In the case of cubic $A$SnI$_3$ ($A$=Rb, Cs), as shown in the Fig.~S4, CsSnI$_3$ has harder low-frequency phonons, and after replacing $|\Phi_2(\rm{Cs-I})|$ by $|\Phi_2(\rm{Rb-I})|$, its low-frequency phonon dispersion is comparable to that of RbSnI$_3$ as seen in Fig.~S4(a, b). However, low-frequency phonons are not sensitive to changes in $A^+$ mass (Fig.~S4(c)). We attribute the phonon dispersion sensitivity to the interaction of $A^+$ cations and halide anions to that the adjacent octahedral structure interacts weakly in metal halide perovskites. The weak interaction makes the low-frequency phonon branch of the octahedral contribution significantly lower in frequency than the optical branch of the $A^+$ contribution, thus preventing coupling together to produce avoid-crossing phenomenon. Since the octahedral structure is mainly connected by ionic bonds, the overlap of electron orbitals between adjacent octahedral atoms is weak as seen in the electronic localization function (ELF) shown in Fig.~S4(d). From the two vibrational modes of the low-frequency phonons (Fig.~S4(e)), it can be seen that the adjacent octahedra rotate independently around their respective centers. The shared halide atoms are not sufficient to significantly enhance the interaction of the neighboring octahedra. The phonon dispersion of double perovskite $A_2$AgBiBr$_6$ ($A$=Rb, Cs) also shows a similar variation as seen in Fig.~S5.

\section*{Conclusions}

In summary, we comprehensively investigated anomalous heat transport behavior in metal halide perovskites $A_2$SnI$_6$ ($A$=K, Rb, Cs). Cs$_2$SnI$_6$ has the largest total atomic mass but the lowest particlelike thermal conductivity $\kappa_p$. The large $\kappa_p$ of Cs$_2$SnI$_6$ results from strong $A$-I interaction, bringing about hardened low-frequency phonon branches and weak lattice anharmonicity. The hardening of low-frequency phonon branches further weakens the phonon scattering phase space and leads to large phonon lifetime. The low-frequency phonon branches are not sensitive to the mass of $A^+$ cations but are sensitive to the interaction between $A^+$ and I$^-$ in [SnI$_6$]$^{2-}$ octahedra, which is inconsistent with the law given by the classical rattling model. This difference comes from the fact that weak interactions between [SnI$_6$]$^{2-}$ octahedra fail to satisfy the condition of strong connections between host frames in the rattling model. Using the established spring model, we have successfully explained the variation rule of the phonon dispersion with $A^+$ cations in $A_2$SnI$_6$ perovskites. Our results are important for understanding the heat transport mechanism of perovskites and designing new materials with low thermal conductivity. 

\section*{Numerical methods}

The calculations are implemented using the Vienna Ab Initio simulation package (VASP) based on density functional theory (DFT)\cite{Kresse1996} with the projector augmented wave (PAW) method and revised PBE-GGA exchange−correlation functional for solids (PBEsol)\cite{Perdew2008}. The cutoff energy of the plane wave is set to 500 eV. The energy convergence value between two consecutive steps is set as $10^{-5}\;$eV when optimizing atomic positions and the maximum Hellmann-Feynman (HF) force acting on each atom is $10^{-3}\;$eV/\r{A}. The calculations of $\kappa_L$ and other relevant parameters such as phonon relaxation time are carried out by the ShengBTE software\cite{shengbte2014,Han2022} which operates based on the iterative scheme. The $\bf{q}$-mesh in the first irreducible Brillouin Zone is set to be 12$\times$12$\times$12. The Gaussian smearing is set as 0.5. A maximum likelihood estimation method is used to accelerate the calculation of the 4ph scattering rate\cite{Guo2024}. The sample size is set as $4\times10^{5}$. The recently introduced on-the-fly Machine Learning Potential (FMLP) of VASP is used to accelerate AIMD simulation. \textcolor{black}{The hyperparameters are set to the default values provided by the VASP and the cutoff radius is 8 \AA, which allows the root mean square error of the energies to be less than $10^{-4}$ eV/atom.} The supercells of 3$\times$3$\times$3 containing 243 atoms are chosen. The simulation is run for 20 ps with a timestep of 1 fs. The second-order, third-order, and fourth-order IFCs are determined from the AIMD simulation by using the TDEP method\cite{Hellman2013}. The cutoff radius of third-order and fourth-order IFCs is set as 6 \AA$\;$and 4 \AA$\;$ respectively.

\section*{Conflicts of interest}
There are no conflicts to declare.

\section*{Acknowledgements}

This work is supported by the Natural Science Foundation of China (12304038), the Startup funds of Outstanding Talents of UESTC (A1098531023601205), National Youth Talents Plan of China (G05QNQR049).

%\section*{Author contributions statement}
%A.A. conceived the experiment(s),  A.A. and B.A. conducted the experiment(s), C.A. and D.A. analysed the results.  All authors reviewed the manuscript.
%\begin{equation}
%\kappa_L^{\alpha \beta}=\kappa_{\mathrm{p}}^{\alpha \beta}+\kappa_{\mathrm{c}}^{\alpha \beta}
%\end{equation}

\providecommand{\latin}[1]{#1}
\makeatletter
\providecommand{\doi}
  {\begingroup\let\do\@makeother\dospecials
  \catcode`\{=1 \catcode`\}=2 \doi@aux}
\providecommand{\doi@aux}[1]{\endgroup\texttt{#1}}
\makeatother
\providecommand*\mcitethebibliography{\thebibliography}
\csname @ifundefined\endcsname{endmcitethebibliography}
  {\let\endmcitethebibliography\endthebibliography}{}

%\bibliography{new1}

\end{document}